\documentclass[prl,floatfix,twocolumn,nobalancelastpage]{revtex4-1}

\usepackage{graphicx}
\usepackage{braket}
\usepackage{amsmath}
\usepackage{amsthm,amsmath,amsfonts,dsfont}

\usepackage{bm}
\usepackage{color}

\newcommand{\comment}[1]{}

\newcommand{\eref}{Eq.~\eqref}

\begin{document}

\title{Quantum many-body dynamics of driven-dissipative Rydberg polaritons}

\author{Tim Pistorius}
\email{tim.pistorius@itp.uni-hannover.de}
\affiliation{Institut f\"ur Theoretische Physik, Leibniz Universit\"at Hannover, Appelstra{\ss}e 2, 30167 Hannover, Germany}
\author{Javad Kazemi}
\affiliation{Institut f\"ur Theoretische Physik, Leibniz Universit\"at Hannover, Appelstra{\ss}e 2, 30167 Hannover, Germany}
\author{Hendrik Weimer}
\affiliation{Institut f\"ur Theoretische Physik, Leibniz Universit\"at Hannover, Appelstra{\ss}e 2, 30167 Hannover, Germany}

\begin{abstract}

  We study the propagation of strongly interacting Rydberg polaritons
  through an atomic medium in a one-dimensional optical lattice. We
  derive an effective single-band Hubbard model to describe the
  dynamics of the dark state polaritons under realistic
  assumptions. Within this model, we analyze the driven-dissipative
  transport of polaritons through the system by considering a coherent
  drive on one side and by including the spontaneous emission of the
  metastable Rydberg state. Using a variational approch to solve the
  many-body problem, we find strong antibunching of the outgoing
  photons despite the losses from the Rydberg state decay.

\end{abstract}


\maketitle


The interplay between external driving and dissipation in strongly
interacting quantum many-body systems leads to the emergence of rich
nonequilibrium dynamics not found in closed quantum systems
\cite{Muller2012, Sieberer2016}, yet their theoretical analysis is
extremely difficult \cite{Weimer2019}. This is especially true in
Rydberg polariton systems \cite{Friedler2005,Dudin2012,Peyronel2012,Firstenberg2013,Cantu2019,Baur2014,Gorniaczyk2014,Tiarks2014,Tiarks2019,Gorshkov2011,Otterbach2013,Bienias2014,Gullans2016}, where the
metastable character of the Rydberg excitation provides a natural
dissipative element. Here, we show that a variational analysis can
successfully describe this challenging many-body problem.

Strongly interacting Rydberg polaritons are closely linked to the
appearance of Electromagnetically Induced Transparency (EIT) involving
a highly excited Rydberg state
\cite{Fleischhauer2000,Fleischhauer2005}. Early experiments have
observed a decline of the EIT feature due to strong Rydberg
interactions \cite{Pritchard2010,Schempp2010}. More recent experiments have
demonstrated the appearance of a strongly interacting polariton
quasi-particle consisting of both light and atomic matter, in a
many-body setting \cite{Peyronel2012,Firstenberg2013,Cantu2019} as well as on the single
polariton level \cite{Baur2014,Gorniaczyk2014,Tiarks2014,Tiarks2019}.  The theoretical analysis of these systems
have so far been limited to an exact treatment of up to two
interacting Rydberg polaritons \cite{Gorshkov2011}, or to large quantum
many-body simulations in the absence of the decay of the Rydberg state
\cite{Otterbach2013,Bienias2014,Gullans2016}.

In this Letter, we investigate the driven-dissipative quantum
many-body dynamics of Rydberg polaritons in an optical lattice
potential. We derive the dispersion relations for the single particle
problem, from which we obtain an effective Bose-Hubbard model for the
dark state polaritons with long-range hopping and long-range
interactions arising from the van der Waals interaction of the Rydberg
states. We show that under experimentally realistic conditions, the
dynamics is confined to a single dark state polariton band, even in
the presence of dissipation from the decay of the Rydberg state and
conversion of dark state polaritons into bright polartions by the van
der Waals interaction. We analyse the driven-dissipative many-body
model using a variational approach, which we
benchmark against wave-function Monte-Carlo simulations for small
system sizes. Finally, we show that strongly correlated photons can be
observed when the polaritons are leaving the system. 


%
%
%
\begin{figure}[thbp]
\includegraphics[width=8.6cm]{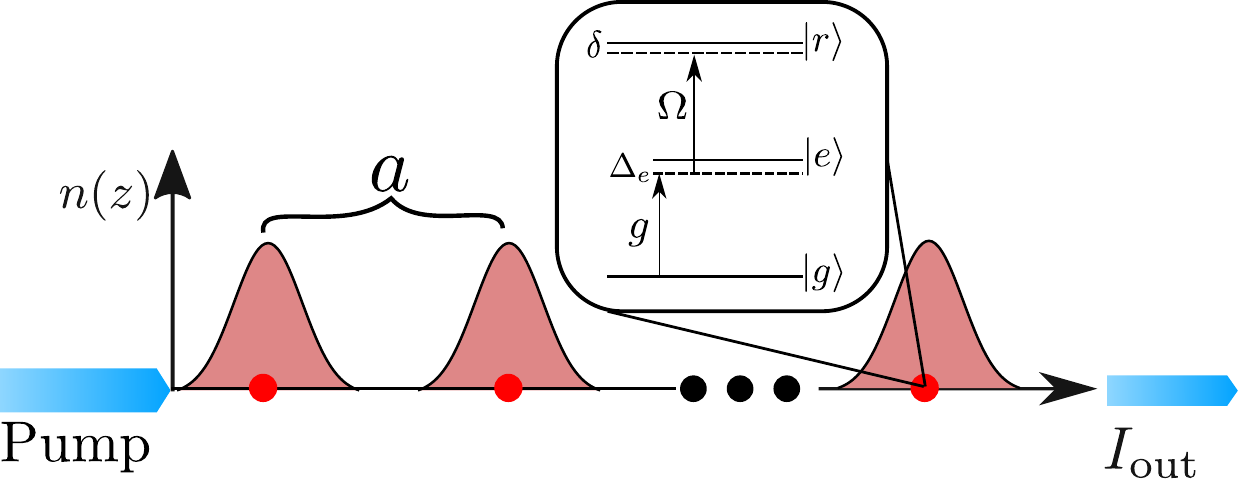} 
\caption{Setup of the system for dark-state polariton propagation. A
  one-dimensional optical lattice potential creates lattice sites
  separated by a distance $a$, around which the atoms exhibit
  approximately Gaussian density profile. The system is being pumped
  from the left by a coherent light field, leading to an output
  intensity $I_\text{out}$.  Each atoms is driven by a photon field
  with a space-dependent coupling $g$ and a coherent laser field
  $\Omega$ with a two-photon detuning $\delta$. The photon field is
  detuned by $\Delta_e$ from the intermediate state.}
\label{fig:system}
\end{figure}
We consider multiple ensembles of rubidium atoms in an effective one-dimensional (1D) geometry with length $L=Na$ with $a$ being the spacing between the $N$ lattice sites created by an appropriate optical lattice potential \cite{Bloch2008}. The atomic density $n(z)$ on each site is approximated by a Gaussian distribution with an average density of $n_0 =10^{13}\ \text{cm}^{-3}$ and a standard deviation of $\sigma=25\,\text{nm}$. Two counterpropagating light fields $\Psi_{E_+}, \Psi_{E_-}$ with the same polarization couple the ground state $\ket{g}$ to a single excited state $\ket{e}$ with a transition frequency of $\omega_{ge}$. The propagation in opposite directions allows for a description in terms of localized Wannier functions \cite{Zimmer2008}. The light fields can be detuned by $\delta_e$ from the atomic transition which we combine with the linewidth $\gamma_e$ of $\ket{e}$ to a complex detuning $\Delta=\delta_e-i\gamma_e$.
A second (control) field with Rabi frequency $\Omega$ enables the transition to a Rydberg state $\ket{r}$ and is set to satisfy a two-photon resonance ($\delta=0$) which brings our system into the EIT regime.
The collective, single-photon Rabi frequency $g(z)$ in this regime is then given by 
\begin{equation} 
g(z)=\tilde{g}\sqrt{n(z)}\sum_l e^{i\tilde{k}la}
\label{eq:coupling}
\end{equation}
 with $\tilde{g}=[6\pi \gamma_e c^3/\omega_{ge}^2]^{1/2}$ and $c$ being the speed of light \cite{Gullans2016}. We split the phase factor up in two parts by setting $\tilde{k}=k_0+k$ which corresponds to the wave vector $k_0=\omega_{ge}/c$ and a deviation from the EIT condition $k$. The transition processes within the atoms can then be described by the bosonic field operators $\hat{\Psi}_p=\ket{g}\bra{e}$ and $\hat{\Psi}_r=\ket{g}\bra{r}$ \cite{Bienias2014}. In the continuum, the non-interacting part of the Hamiltonian can then be written as 
\begin{equation}
H_0=\hbar \int dz\ \boldsymbol{\hat{\Psi}}^\dagger
\begin{pmatrix} -ic\partial_z &0& g(z) &0 \\ 0 &ic\partial_z& g(z) &0 \\g(z) &g(z)& \Delta &\Omega \\ 0 &0& \Omega &\delta  \end{pmatrix} \boldsymbol{\hat{\Psi}}.
\label{eq: start_hamiltonian}
\end{equation}
with $\boldsymbol{\hat{\Psi}}=\{\hat{\Psi}_{E_+}, \hat{\Psi}_{E_-}, \hat{\Psi}_{e}, \hat{\Psi}_{r}\}$. The kinetic terms for the quantized light fields only account for the previous mentioned deviation from the two-photon resonance.\\

We obtain the single polariton solution of \eref{eq: start_hamiltonian} by using a Bloch wave ansatz $\bm{\phi}_k(z)=e^{ikz} \mathbf{u}_k(z)$ in combination with a plane wave expansion for the periodic functions $\mathbf{u}_k(z)$.
The eigenstates of the resulting band structure are a composition of the previously defined bosonic fields and can be interpreted as polaritons \cite{Fleischhauer2002}. Most eigenstates will dissipate quickly because of the spontaneous emission rate that arises from any contribution of $\ket{e}$. Hence, we want to focus on the dark-state polaritons with their vanishing population $\braket{\hat{\Psi}_{e}^\dagger\hat{\Psi}_{e}}$.

\begin{figure}[thbp]
\includegraphics[width=8.6cm]{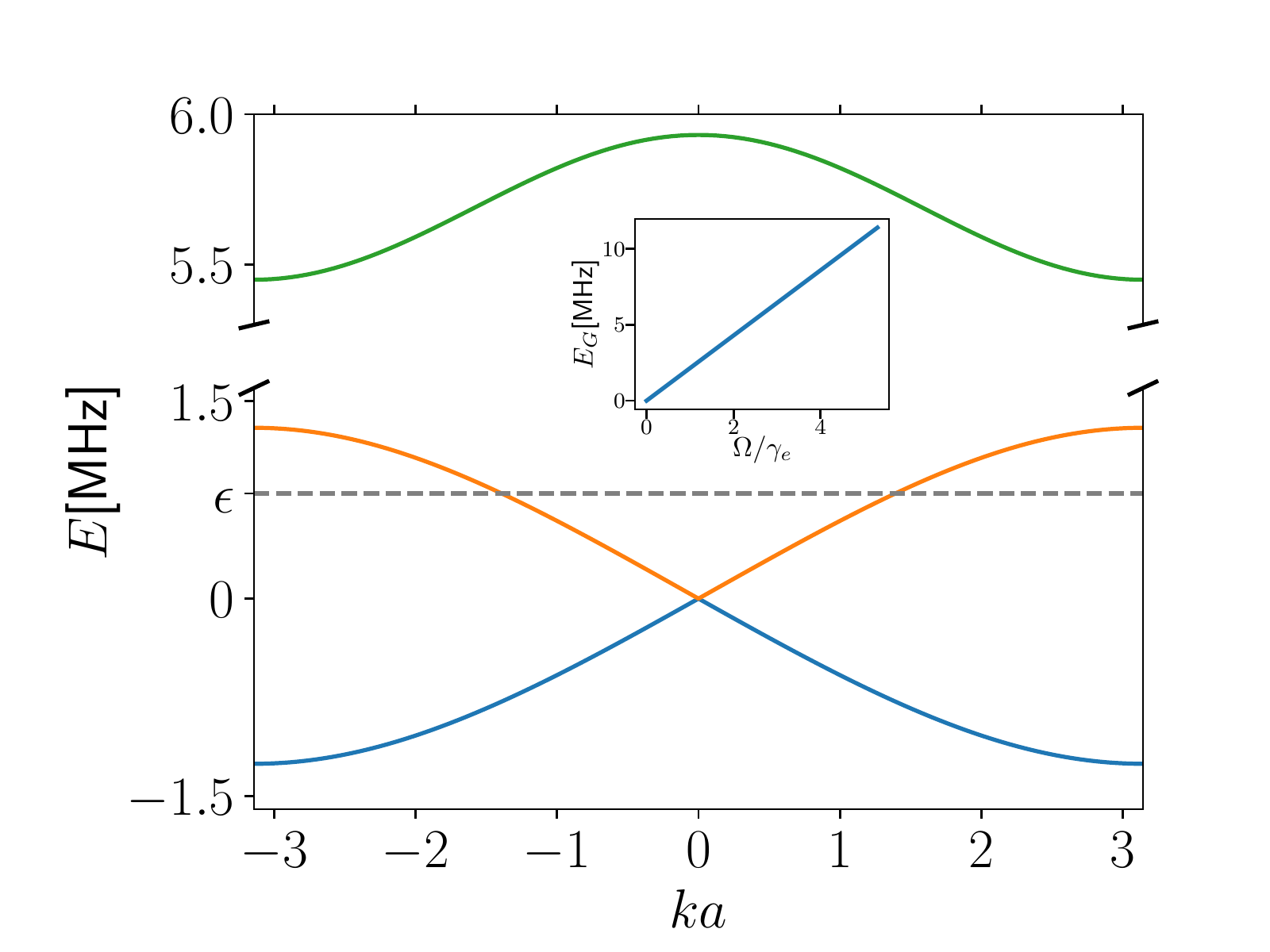} 
\caption{Dispersion relation for polaritons close to zero energy for $\Omega/2\pi=18\ \text{MHz}$, $\delta_e/2\pi=20\ \text{MHz}$, $\gamma_e/2\pi=6\ \text{MHz}$, and $a=532 \text{nm}$. We obtain two dark-state polariton bands and exemplarily show a bright state polariton band. The dashed grey line indicates the average energy $\epsilon$ of the upper dark state polariton. The insert shows the scaling of the band gap with $\Omega$. }
\label{fig:dispersion}
\end{figure}

The lower part of Fig. \ref{fig:dispersion} shows their dispersion relations for a typical excitation scheme $5s\rightarrow 5p\rightarrow 34s_{1/2}$ in ${}^{87}$Rb. The coupling of the forward and backward propagating light field to the same intermediate level leads to a symmetric behaviour of the bands and results in a linear dispersion. The solution at $k=0$ presents a superposition of both bands which results in a cancellation of the Rydberg part in the polaritons and a crossing of the bands at that point \cite{Iakoupov2016}.
The surrounding bands like the one shown in the upper part of Fig. \ref{fig:dispersion} are separated by a large band gap compared to the energy scale of the dynamics of the dark states, resulting in the dynamics confined to the bands close to zero energy.


In the following, we transform the eigenstates of \eref{eq: start_hamiltonian} into localized Wannier functions $\mathbf{w}_j(z)=\frac{1}{\sqrt{N}}\sum_{k} e^{-ikaj}\bm{\phi}_k(z)$, resulting in bosonic creation operators $a_i^\dagger = \int dz \mathbf{w}(z)\mathbf{\Psi}(z)$ for the upper band and the analogous operators $b_i$ for the lower band. Additionally, we consider a pumping term $P$ on the first lattice site, describing the driving with a coherent light field from the left. In the Wannier basis, the Hamiltonian has the form
\begin{align}
H_0&= -\sum_{i,j} J_{i,j} (\hat{a}_i^\dagger \hat{a}_j-\hat{b}_i^\dagger \hat{b}_j+h.c.) \nonumber \\
&+(2\epsilon-\beta)\sum_i \hat{b}_i^\dagger \hat{b}_i +\beta \sum_i \hat{a}_i^\dagger \hat{a}_i \label{eq:tp_bh} \\
&+p (\hat{a}_1^\dagger+\hat{a}_1+\hat{b}_1^\dagger+\hat{b}_1). \nonumber
\end{align}
The first line in Eq. \eqref{eq:tp_bh} describes hopping between the sites with a strength of $J_{i,j}$, which can be written in terms of the hopping length $m$ as $J_m$ with $|i-j|=ma$. It is important to note that the scaling of $J_{i,j}$ with the distance $|i-j|$ does not follow an exponential decay but a power law asymptotically decaying like $|i-j|^{-2}$, which arises from the linear dispersion of the bands at around $k\approx 0$. Hence, we cannot approximate the system by a nearest neighbor-hopping $J_1$, which is possible when considering different level schemes \cite{Fleischhauer2004}.
 The following two terms are the on-site energy shifts where the factor $\beta$ indicates the detuning from a resonant driving of the upper polariton branch. For $J_1/(2\epsilon-\beta)\ll 1$ the lower band is far detuned and can be neglected.

 Let us now consider the consequences of the van der Waals interaction
 $V(z)=C_6/z^6$ between atoms in the Rydberg state on our system to
 see if the assumptions we made so far still hold true. The repulsive
 nature of the interaction leads to a blockade radius inside the
 lattice which is defined through the strength of the hopping $J_{1}$
 between different sites and the van der Waals coefficient $C_6$ for
 the chosen Rydberg state $\tilde{r}_{b}=\sqrt[6]{\frac{C_6}{J_{1}}}$,
 similar to the conventional Rydberg blockade for stationary atoms
 \cite{Jaksch2000}. An important consequence of the van der Waals
 force is the two-photon detuning for atoms in the vicinity of an
 already excited atom which exceeds the EIT linewidth of the system at
 a characteristic distance $r_{b}=\sqrt[6]{\frac{C_6
     |\Delta|}{\Omega^2}}$. Below that distance the EIT window brakes
 and photons can get absorbed into the intermediate state of the
 atoms. This causes a scattering of the photons and restricts the
 creation of a new polariton only to sites outside $r_b$. This also
 allows to restrict the pumping term in Eq. \eqref{eq:tp_bh} to the
 first site of the lattice \cite{Zeuthen2017}. In our case where
 $r_b>\tilde{r}_b$ it does not affect the internal many-body dynamics
 between multiple polaritons but it is used as a regularization for
 the calculation of the interaction strength between them which is
 then given by
 \begin{equation}
 V_{ij}= \frac{C_6}{2} \int dz dz' \frac{\mathbf{w}_i^*(z) \mathbf{w}_j^*(z') \mathbf{w}_j(z') \mathbf{w}_i(z)}{r_b^6+|z-z'|^6}.
 \end{equation}
At distances larger than $r_b$, the interaction energy is small compared to the band gap between the dark-state polaritons and the other bands so that the single band approximation still holds true. Also, it restricts the numbers of polaritons on each site to a single excitation which we can implement by choosing Pauli operators $\sigma^{-(+)}$ for the annihaltion(creation) operators $a^{(\dagger)}$ in Eq. \eqref{eq:tp_bh}. 
  Putting everything together gives us an extended Bose-Hubbard Hamiltonian for interacting dark-state polaritons
 \begin{equation}
 \begin{aligned}
 H=& -\sum_{i,j} J_{i,j} \sigma_i^+ \sigma^-_j+P (\sigma^+_1+\sigma^-_1)\\
&+\beta\sum_i \sigma_i^+ \sigma^-_i+\sum_{i,j}V_{ij} \sigma_i^+ \sigma_j^+ \sigma^-_i \sigma^-_j.
 \label{eq:Bose-Hubbard}
 \end{aligned}
 \end{equation}
 
So far we have neglected the second natural dissipation channel in our system in form of the spontaneous decay from the Rydberg state. To describe the dynamics of the open quantum system under the condition of Markovianity we can use the Lindblad form of the differential equation $\frac{d}{dt}\rho=\mathcal{L}\rho$ with the Liouvillian $\mathcal{L}$ being the generator of the dynamics \cite{Breuer2002},i.e.
\begin{equation}
\mathcal{L}(\rho)=-i [H(t),\rho(t)]+\sum_j \left(c_j\rho c_j^\dagger-\frac{1}{2}\{c_j^\dagger c_j,\rho\}\right).
\end{equation}
 
The spontaneous emission from the Rydberg state also effects the polaritons and is described by the jump operators $c_i=\sqrt{\gamma_i} \sigma_-^{(i)}$ for each site $i$ and an effective decay rate of $\gamma_i = \int dz |w_r^{(i)}(z)|^2\gamma_r=12.5\text{kHz}$, with $\gamma_r$ being the decay rate of the Rydberg state \cite{Branden2009}. Additionally, to account for photons leaving the system along the propagation axis we add another dissipation channel with jump operators $c_{1(,N)}=\gamma_\text{out} \sigma^-_{1(,N)}$ that only applies on the first and last site of the lattice. Here, we consider the case where $\gamma_\text{out} = J_1$, i.e., the coupling to the outside has the same strength as the internal nearest-neighbor hopping. Similar processes can also be defined for the other sites but show an insignificant influence on the overall dynamics. This allows us to compute the output photon intensity in means of the internal dynamics of the polaritons in the system, providing a similar approach as the input-output formalism in for example cavity QED systems \cite{Caneva2015,Collett1985}. Here, we can define the output intensity as $I_\text{out}=\kappa J_1\braket{\sigma_N^+ \sigma_N^-}$.  In the following, we will drop the proportionality factor $\kappa$ in the calculations for convenience. 

We perform exact numerical simulations of the system for site
numbers up to $N=10$ via the wave-function Monte-Carlo method
using the QuTiP library \cite{Johansson2013}, which is amounts to an
average of about two polaritons inside the system. In all our
simulations, we choose the initial state to have no polaritons in the
system. To analyze the output for larger lattices, we use a
variational approach \cite{Weimer2015,Overbeck2016} starting with a
product ansatz for the density matrix
\begin{equation}
\rho= \prod_{i = 1}^{N} \rho_{i}=\frac{1}{2} \prod_{i = 1}^{N}\left( 1+\sum_{\mu\in \{x,y,z\}} \alpha_\mu \sigma_\mu^{i}\right)
\end{equation}
with $\rho_i$ as the density matrix for each lattice site and
$\alpha_\mu$ as our variational parameters. This product state is then
restricted to a blockade constraint for the polaritons, such thate
there is only one polariton inside a blockade radius $r_b$, i.e.,
$\sum_{i-r_b<j<i+r_b} \langle \sigma_+^{(i)}\sigma_-^{(i)}\rangle \leq
1$ for all sites $i$. This approach is equivalent to the hard sphere
correlation function used in the analysis of coherently driven Rydberg
gases \cite{Weimer2008a}.

For the variational integration of the quantum master equation, we use
an implicit midpoint method \cite{Overbeck2016}. To reduce the number
of variational parameters in a single optimization, we evolve the
system from $t$ to $t+\Delta t$ by minimizing the parameters for one
site and hold every other site constant \cite{Overbeck2017}. This
procedure is repeated for all sites before moving on to the next time
step. For the variational optimization we use the norm $D_i$ for each site
$i$ given by
\begin{equation}
\begin{split}
D_i=&\sum_{j\neq i} ||-\frac{\tau}{2}\mathcal{L}[\rho_i(t+\tau)\rho_j(t)+\rho_{ij}(t)]\\
&+\rho_i(t+\tau)\rho_j(t)-\rho_{ij}(t)||_1\rightarrow \text{min},
\end{split}
\end{equation}
where $||\cdot ||_1$ denotes the trace norm given by $\text{Tr}\{|\cdot|\}$. Additionally, we add constraints to the minimization to enforce the positivity of the density matrix $\rho_i\geq 0$ and to enforce the blockade of the polaritons.

\begin{figure}[t]
\includegraphics[width=8.6cm]{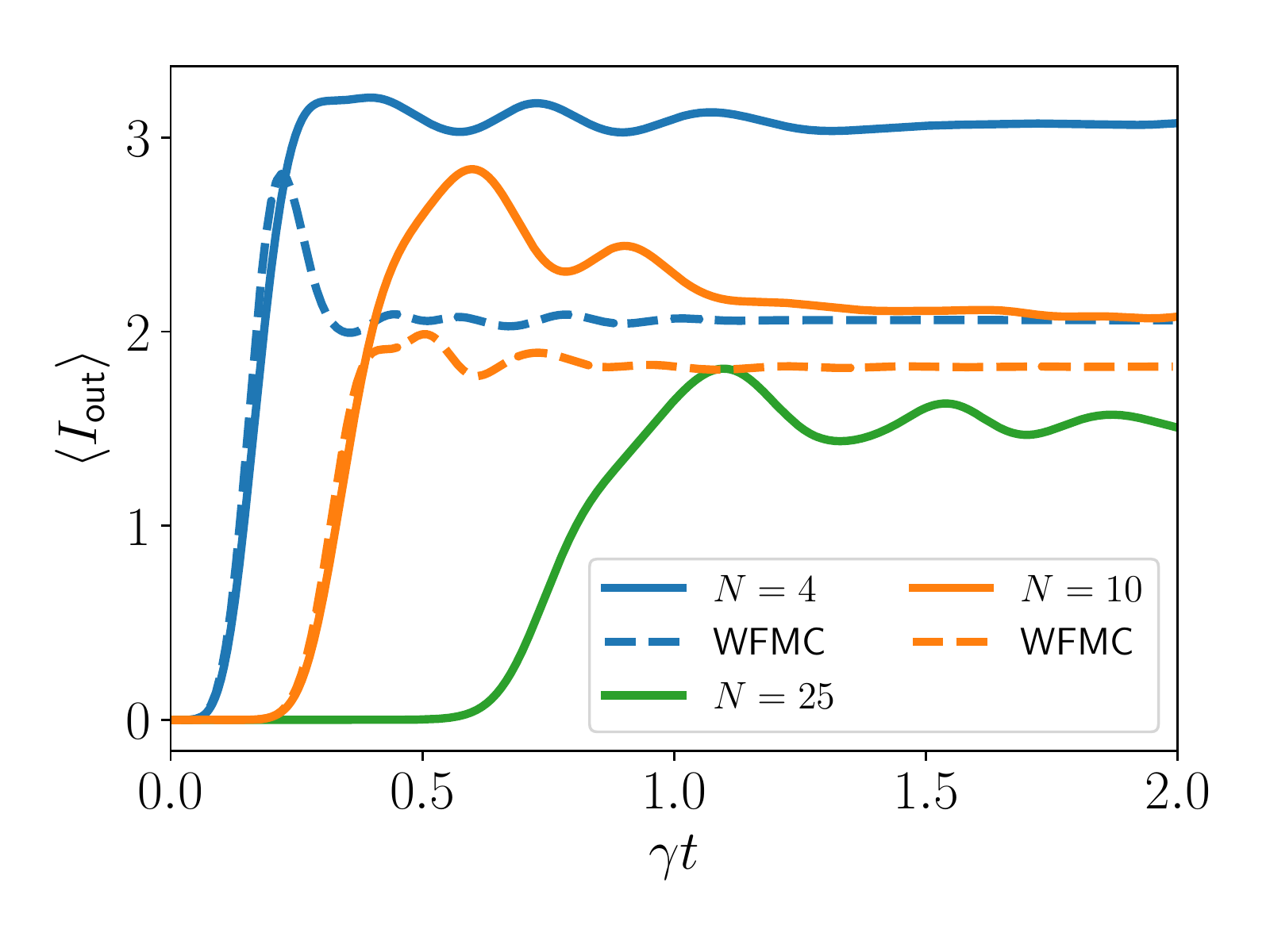}
\caption{Intensity output $\braket{I_\text{out}}$ for different system sizes for a pump strength of $P=10\,\gamma$. For smaller system size ($N=4,10$) the variational approach (solid) is compared to wave function Monte-Carlo (WFMC) simulations (dashed). }
\label{fig:I_out}
\end{figure}
\begin{figure}[t]
\includegraphics[width=8.6cm,trim=0.cm 0cm 0cm 0cm,clip]{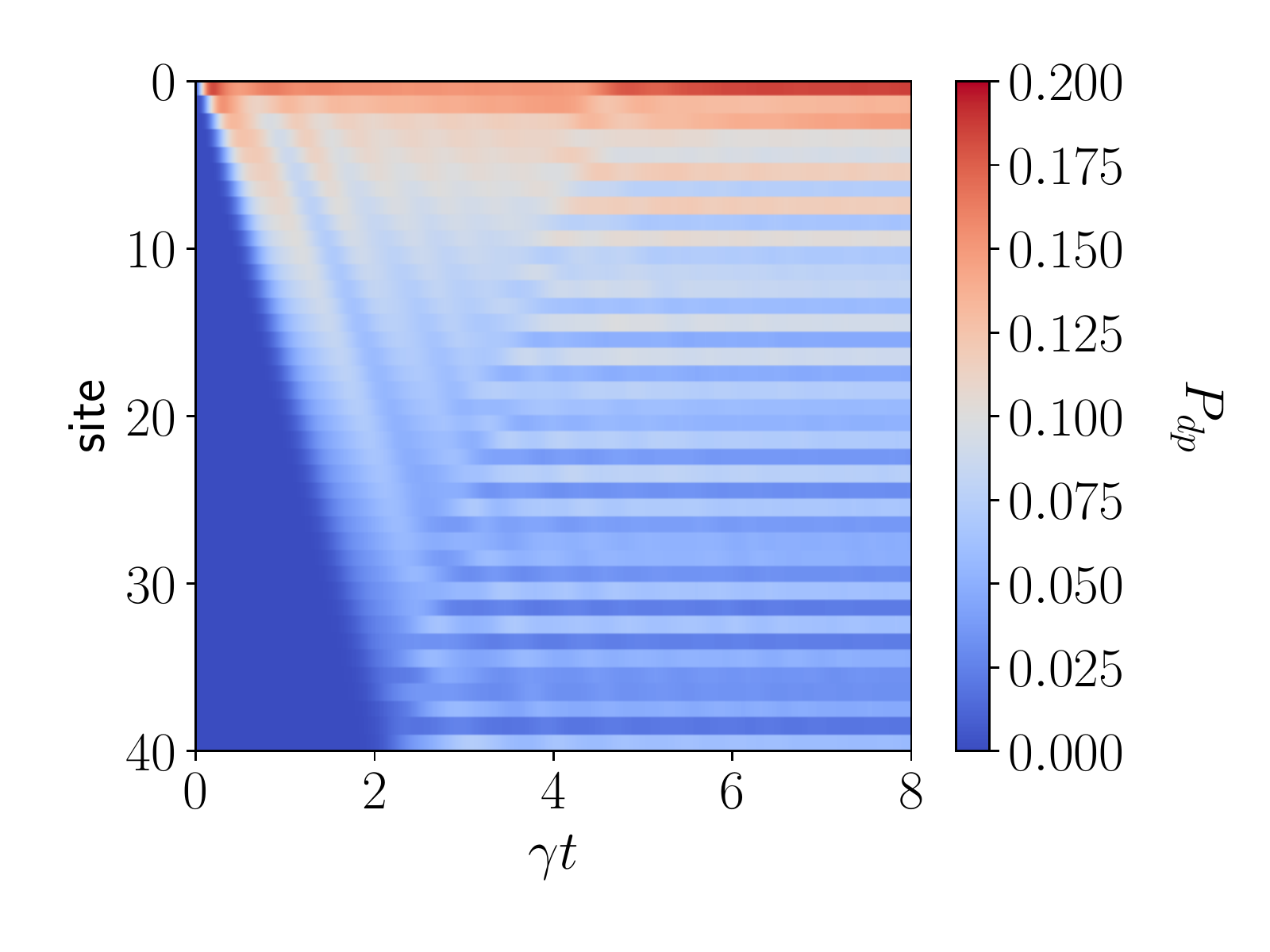} 
\caption{Time evolution of the polariton population $P_{dp}^{(i)}=\langle \sigma_+^{(i)}\sigma_-^{(i)}\rangle $ of each site $i$ in a lattice of size $N=40$ for a pumping strength of $P=10\,\gamma$. }

\label{fig:I_out_all_sites}

\end{figure}

 
Figure \ref{fig:I_out} shows the intensity output $I_\text{out}$ for different lattices sizes $N$. Additionally, we benchmark the variational results against wave-function Monte-Carlo simulations. We find that the two are in good agreement, especially for larger system sizes.
Having demonstrated the viability of the variational approach, we now
turn to the variational simulation of larger system sizes.  Figure
\ref{fig:I_out_all_sites} displays the dynamics of the polariton
population on each site for a lattice size of $N=40$. We observe that
a significant portion of the polariton density remains confined to the
initial pump site, with the rest of the population spreading
throughout the system similar to a light cone, which is a consequence
of the linear dispersion relation.

Finally, we also want to look at the temporal correlations in the
output intensity. For this, we let the system evolve until it reaches
a steady state at time $t_\text{ss}$. At this time, we consider the effect of a quantum jump corresponding to a photon leaving the system, after which we let the system evolve for an additional time $\tau$. Then, the probability to observe a second photon is described by the two-time correlation function

\begin{align}
g^{(2)}(\tau)&=\frac{\braket{\sigma_N^+(t_\text{ss})\sigma_N^+(t_\text{ss}+\tau)\sigma_N^-(t_\text{ss}+\tau)\sigma_N^-(t_\text{ss})}}{\braket{\sigma_N^+ \sigma_N^-}_ {t_\text{ss}}^2} \nonumber\\
&=\frac{1}{\braket{\sigma_N^+ \sigma_N^-}_ {\text{ss}}^2}  \text{Tr}\left\{\sigma_N^- e^{\mathcal{L}\tau}\left[\sigma_N^- \rho(t_\text{ss})\sigma_N^+\right]\sigma_N^+\right\},
\label{eq:time_correlation}
\end{align} 
where we have used the cyclicity of the trace \cite{Zoller2000}. The first measurement results in a complete setback for the polariton excitation probability on all sites inside the blockade radius from the last site. We use a self-consistent approach to identify this distance by adding the excitation probabilities of the other sites beginning from site $N-1$ until $\sum_{i=N-1} \braket{\sigma_i^+ \sigma_i^-}=1$ and set them back to the ground state. The blockaded region which is defined in that way is for smaller system size identical to our previous definition of the blockade radius. For larger system sizes, the radius is extended because it takes the consequences of the decay from the Rydberg state into account.
\begin{figure}[b]
\includegraphics[width=8.6cm]{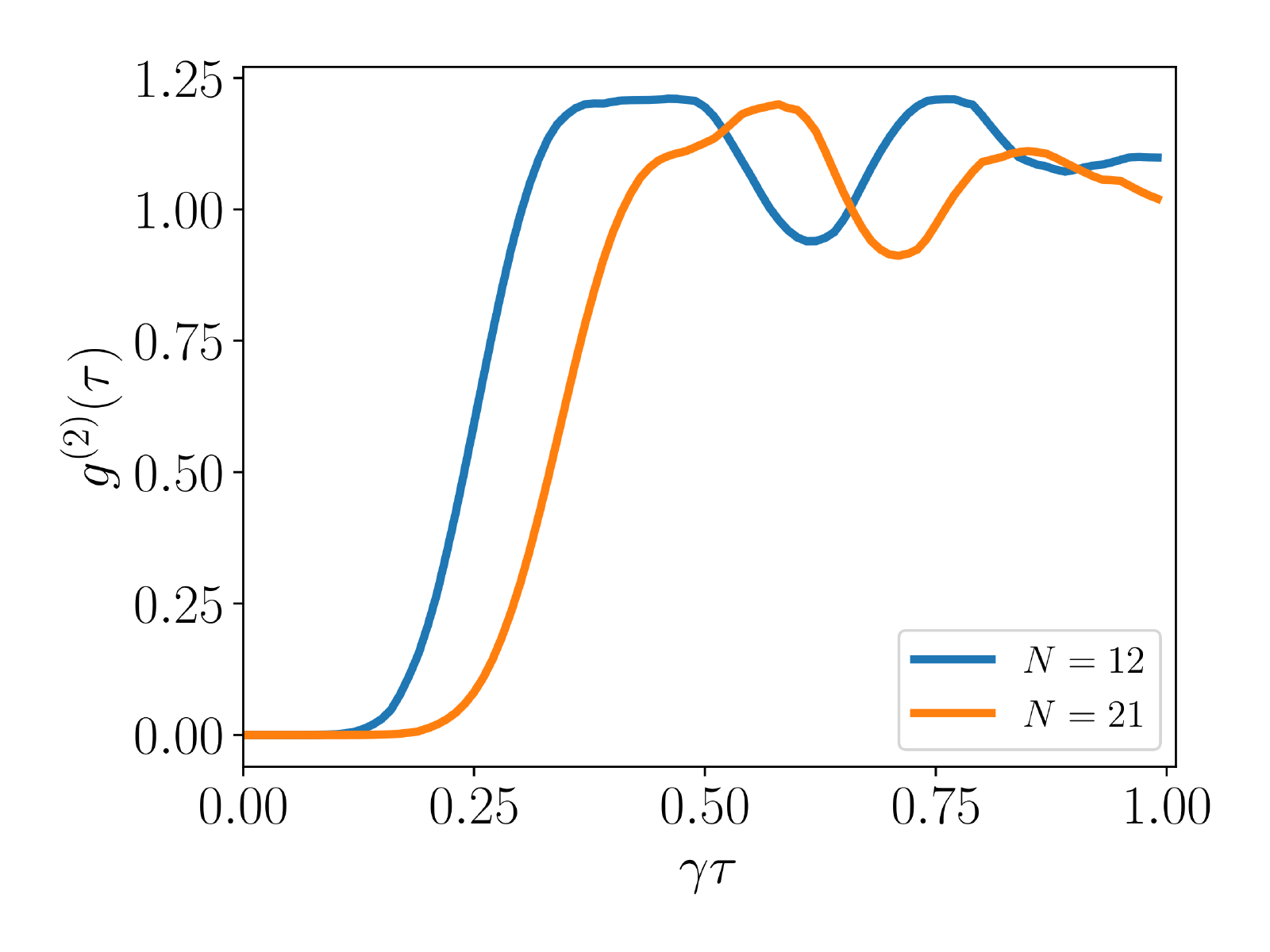}
\caption{Two time correlation function $g^{(2)}(\tau)$ of the output signal from the last site of the lattice for different system sizes $N$. }
\label{fig:time correlation}
\end{figure}

Figure \ref{fig:time correlation} shows an extended anti-bunched
region [$g^{(2)}(\tau)\approx 0$] resulting from the blockade. At
later times, We also observe bunching before the system goes back to
the steady-state value of $g^{(2)}(\tau)=1$. These findings underline
the possibility of using Rydberg polariton systems to generate
strongly correlated photon streams, similar as it has been discussed
for free-space systems \cite{Zeuthen2017}.

In summary, we have demonstrated the possibility to treat large
many-body systems of driven-dissipative systems of strongly
interacting Rydberg polaritons using a variational approach. Deriving
an extended Bose-Hubbard model with long-range hopping and
interactions, we observe that the propagation of photons through a
lattice can yield in strong correlations between the particles. The
variational approach proved to be a good approximation for the
dynamics especially for larger system sizes.  Our work presents a
first look into the driven-dissipative transport of Rydberg polaritons
and paves the way for future investigations of different driving
scenarios and extensions to free-space polaritons in the form of a
suitable continuum limit.


\begin{acknowledgments}
We thank H.P.~B\"uchler for fruitful discussions. This work was funded
by the Volkswagen Foundation, by the Deutsche Forschungsgemeinschaft
(DFG, German Research Foundation) within SFB 1227 (DQ-mat, project
A04), SPP 1929 (GiRyd), and under Germany’s Excellence Strategy --
EXC-2123 QuantumFrontiers -- 390837967.
\end{acknowledgments}

\bibliography{bib_polariton_paper}


\end{document}